\documentclass[lettersize,journal]{IEEEtran}
\usepackage{amsmath,amsfonts}
\usepackage{algorithmic}
\usepackage{algorithm}
\usepackage{array}
\usepackage[caption=false,font=normalsize,labelfont=sf,textfont=sf]{subfig}
\usepackage{textcomp}
\usepackage{stfloats}
\usepackage{url}
\usepackage{verbatim}
\usepackage{graphicx}
\usepackage{cite}

\usepackage[table]{xcolor}
\usepackage{colortbl} 
\usepackage{tabularx}
\usepackage{amsmath} 
\usepackage{wasysym}
\usepackage{arydshln}
\definecolor{lightblue}{rgb}{0.85, 0.95, 1.0}
\usepackage{times}
\usepackage{helvet}
\usepackage{courier}
\usepackage{pifont}
\usepackage{multirow}
\usepackage{booktabs}
\usepackage{orcidlink}
\usepackage{enumitem}
\usepackage[most]{tcolorbox}
\tcbuselibrary{listings, breakable, skins}
\usepackage{hyperref}
\usepackage{url}

\newtcolorbox{promptbox*}[1][]{
    colback=gray!10,
    colframe=gray!50,
    fontupper=\small\ttfamily,
    breakable=unlimited,
    enhanced,
    width=\linewidth,
    title=Prompt Example,
    titlerule=0pt,        
    boxrule=0.5pt,        
    frame hidden=false,   
    break at=\textheight, 
    #1                   
}

\begin{document}

\title{DegDiT: Controllable Audio Generation with Dynamic Event Graph Guided Diffusion Transformer}

\author{Yisu Liu\textsuperscript{*}\orcidlink{0009-0002-8490-472X}, Chenxing Li\textsuperscript{*}\orcidlink{0000-0002-6644-5104}, 
Wanqian Zhang\textsuperscript{\textdagger}\orcidlink{0000-0001-5734-4072}, Wenfu Wang, Meng Yu, Ruibo Fu\orcidlink{0000-0001-9598-1881}, \\Zheng Lin\orcidlink{0000-0002-8432-1658}, Weiping Wang\orcidlink{0000-0002-8618-4992}, and Dong Yu\textsuperscript{\textdagger}\orcidlink{0000-0003-0520-6844},~\IEEEmembership{Fellow,~IEEE}
\thanks{\textsuperscript{*} denotes equal contribution. \textsuperscript{\textdagger} denotes corresponding author.
}
\thanks{Yisu Liu, Zheng Lin, and Weiping Wang are with the Institute of Information Engineering, Chinese Academy of Sciences, Beijing 100085, China, and also with the School of Cyber Security, University of Chinese Academy of Sciences, Beijing 101408, China. (E-mail: liuyisu@iie.ac.cn, linzheng@iie.ac.cn, wangweiping@iie.ac.cn)}
\thanks{Chenxing Li and Wenfu Wang are with the Tencent, AI Lab, Beijing 100089, China (E-mail: chenxingli@tencent.com, wenfuwang@tencent.com).}
\thanks{Wanqian Zhang is with the Institute of Information Engineering, Chinese Academy of Sciences, Beijing 100085, China. (E-mail: zhangwanqian@iie.ac.cn)}
\thanks{Ruibo Fu is with the Institute of Automation, Chinese Academy of Sciences, Beijing 100089, China (E-mail: ruibofu@126.com).}
\thanks{Meng Yu and Dong Yu are with the Tencent, AI Lab, Bellevue, WA 98004, USA (E-mail:raymondmyu@global.tencent.com, 
dyu@global.tencent.com).}
}

\markboth{Journal of \LaTeX\ Class Files,~Vol.~14, No.~8, August~2021}%
{Shell \MakeLowercase{\textit{et al.}}: A Sample Article Using IEEEtran.cls for IEEE Journals}

\maketitle
\begin{abstract}
Controllable text-to-audio generation aims to synthesize audio from textual descriptions while satisfying user-specified constraints, including event types, onset and offset timestamps, and temporal sequences. 
This enables precise control over both the content and temporal structure of the generated audio. 
Despite recent progress, existing methods still face inherent trade-offs among accurate temporal localization, open-vocabulary scalability, and practical efficiency.
To address these challenges, we propose \textbf{DegDiT}, a novel dynamic event graph-guided diffusion transformer framework for open-vocabulary controllable audio generation. 
DegDiT encodes the events from the text description into structured dynamic graphs. 
The nodes represent distinct audio events, while edges encode the temporal relationships between them.
A transformer is employed to integrate these graphs and produce contextualized event embeddings that serve as guidance for the diffusion model. 
To ensure high-quality and diverse training data, we introduce a quality-balanced data selection pipeline that combines hierarchical event annotation with multi-criteria quality scoring, resulting in a curated dataset with semantic diversity.
Furthermore, we present consensus preference optimization, facilitating audio generation through consensus among multiple reward signals.
Extensive experiments on AudioCondition, DESED, and AudioTime datasets demonstrate that DegDiT achieves state-of-the-art performances across a variety of objective and subjective evaluation metrics.
Some generated examples are available at \href{https://riolys.github.io/DegDiT}{\textcolor{blue}{DegDiT Demo Page}}.
\end{abstract}

\section{Introduction}
Diffusion models \cite{croitoru2023diffusion, rombach2022high} have established themselves as a dominant framework for generative artificial intelligence, demonstrating remarkable success across various modalities.
These models, grounded in the theory of stochastic differential equations \cite{songscore}, gradually transform the random noise into high-quality samples.
This iterative refinement process has demonstrated particular effectiveness for complex data distributions, leading to breakthroughs in image generation with models like Stable Diffusion \cite{stabilityai} and DALL·E \cite{flux}, as well as in video synthesis such as Sora \cite{sora2024}.
The inherent flexibility of diffusion models has also enabled their successful adaptation to audio generation tasks \cite{kreuk2022audiogen}, where they have shown superior performance compared to traditional generative approaches like GANs \cite{goodfellow2014generative,karras2020analyzing} or autoregressive models \cite{xiong2024autoregressive}.

\begin{figure}
\centering 
\includegraphics[width=0.47\textwidth]{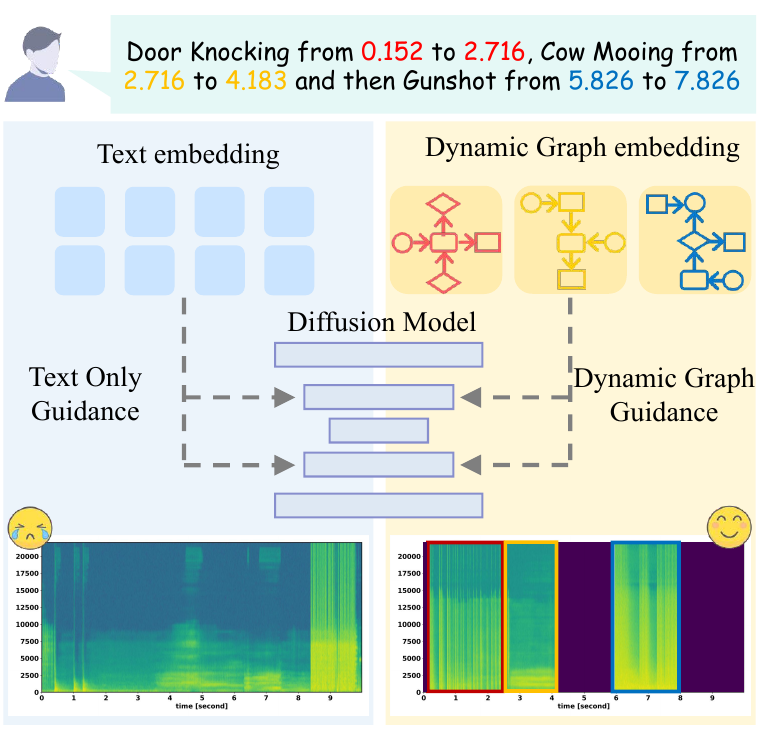}
\caption{Comparison between naive text-only guided (left) and our proposed dynamic event graph guided (right) diffusion model generation.
The naive text-only guided approach relies solely on textual input to drive the diffusion process.
By integrating structured event graphs, our approach effectively captures temporal dependencies, enabling the generation of accurate and consistent timestamps.} 
\label{Fig1} 
\end{figure}

Building upon these advances, text-to-audio (TTA) generation \cite{huang2023make, ghosal2023text, xue2024auffusion} has emerged as a particularly promising application of diffusion models.
TTA systems aim to synthesize realistic audio content from natural language descriptions, offering transformative potential for multimedia production, accessibility tools, and creative applications.
Current state-of-the-art systems like AudioLDM \cite{liu2023audioldm, audioldm2-2024taslp}, Tango \cite{ghosal2023text,majumder2024tango} and TangoFlux \cite{hung2024tangoflux} have demonstrated impressive capabilities in audio generation.
Despite these advances, achieving fine-grained control over the generated audio remains a significant challenge for TTA systems.
Such limitations hinder the use of TTA systems in applications that require detailed audio control.
To solve this problem, one might consider drawing inspiration from controllable image generation methods \cite{zhang2023adding, guo2024pulid,yuan2025identity}, which have achieved remarkable success in manipulating visual content according to the user-specified conditions.
However, these approaches cannot be directly adapted to audio due to modality differences, especially the temporal dynamics and continuous nature of audio signals.

Toward enabling precise control of text-to-audio generation, recent works have explored various strategies.
As an early exploration, MC-Diffusion \cite{guo2024audio} introduces multi-scale conditioning to simultaneously govern temporal and spectral attributes of generated audio.
PicoAudio \cite{xie2025picoaudio} addresses the precision of timestamps by leveraging LLM to conduct a class-timestamp matrix as guidance.
However, this approach is limited to a fixed set of classes due to the predefined dimensionality of the matrix.
While AudioComposer \cite{wang2025audiocomposer} introduces a fully end-to-end natural language-based generation approach, it often suffers from missing audio events and degraded output quality due to its over-reliance on text descriptions.
Most recently, FreeAudio \cite{jiang2025freeaudio} introduces a training-free method using aggregation of preliminary and timestamp-independent attention maps, which introduces operational complexity due to its reliance on carefully designed parameters.

To address these issues, we propose DegDiT, which models sound events within the multi-event audio prompt as temporally dynamic graphs that evolve across the audio timeline, as depicted in Fig. \ref{Fig1}.
In the graphs, each event is represented as a node carrying categories, temporal activation, and boundary attributes, while the edges between nodes encode their temporal relationships.
A graph transformer processes these dynamic graphs, aggregating the spatio-temporal context through attention mechanisms to produce contextualized embeddings. 
These enriched embeddings are then fused with the textual features and provided as guidance for the diffusion model.

To establish an open-vocabulary setting while enhancing dataset diversity, we also introduce a Quality-Balanced Data Selection pipeline that combines hierarchical event annotation with multi-criteria quality scoring.
Our approach utilizes HTS-AT \cite{chen2022hts} to detect the event boundaries and generate confidence-calibrated pseudo-labels of AudioSet \cite{audioset} samples.
We then conduct a systematic evaluation of each sample through four key metrics: event count, type diversity, temporal alignment accuracy, and duration plausibility, yielding a rigorously curated dataset.
The diffusion model and the graph transformer are jointly trained on this curated dataset, followed by fine-tuning on the AudioCondition dataset \cite{guo2024audio}.

Considering the multi-dimensional evaluation requirements in controllable audio generation, we propose Consensus Preference Optimization (CoPO), a reinforcement learning framework that integrates diverse reward signals for preference modeling.
Unlike conventional binary preference learning, CoPO leverages quantitative preference intensities derived from quality assessments across multiple dimensions: text alignment, event alignment, temporal accuracy, and audio quality.
By aggregating these reward components with learnable weights, CoPO captures the consensus among multiple reward signals, facilitating more nuanced model optimization.

In brief, our contributions are summarized as:

\begin{enumerate}
    \item We introduce DegDiT, a diffusion transformer framework for open-vocabulary, fine-grained controllable audio generation.
    The temporally dynamic event graphs are designed to encode precise temporal attributes and inter-event relationships, which are modeled using a graph transformer encoder as guidance for the diffusion model.
    \item A Quality-Balanced Data Selection pipeline is proposed to combine hierarchical event annotation and multi-criteria quality scoring, resulting in a curated dataset with acoustic and semantic diversity for model training.
    \item We further develop Consensus Preference Optimization, a reinforcement learning-based method that leverages multi-dimensional reward signals for fine-grained preference modeling.
    \item Extensive experiments show that DegDiT achieves state-of-the-art performance on the AudioCondition \cite{guo2024audio}, DESED \cite{turpault2019desed}, and AudioTime \cite{xie2025audiotime} datasets across various subjective and objective evaluation metrics.
\end{enumerate}

\section{Related Work}
\subsection{Text-to-Audio Generation}
The advances in text-to-audio (TTA) generation have been propelled by the development of diffusion models.
Auffusion \cite{xue2024auffusion} combines the strengths of diffusion models and large language models (LLMs) to improve the quality of TTA generation.
Make-an-Audio \cite{huang2023make} and AudioLDM \cite{liu2023audioldm, audioldm2-2024taslp} have explored audio generation in continuous latent spaces, further improving the fidelity and diversity of generated audio.
To enhance training efficiency, Tango \cite{ghosal2023text,majumder2024tango} utilizes the language understanding capabilities of LLMs to condition the diffusion model.
More recently, EZAudio \cite{hai2024ezaudio} presents an efficient TTA framework building on diffusion transformer architecture.
Meanwhile, TangoFlux \cite{hung2024tangoflux} combines flow matching with CLAP-guided preference optimization, achieving real-time generation while maintaining high audio fidelity. 

\subsection{Controllable Image Generation}
To achieve precise spatial control in the image generation era, researchers employ auxiliary control signals to guide the diffusion process.
ControlNet \cite{zhang2023adding} utilizes structural cues such as depth maps to constrain the generation process, ensuring output images adhere to predefined spatial configurations.
Subsequent works, including IP-Adapter \cite{ye2023ip}, PulID \cite{guo2024pulid}, and ConsisID \cite{yuan2025identity}, have further advanced this approach by incorporating dedicated image encoders to extract spatial features from the reference images.
Beyond these trained adapters, training-free methods have emerged to enhance flexibility and reduce computational overhead.
Regional Prompting \cite{chen2024training} achieves spatial control using region masks in pre-trained models, and RPG \cite{yang2024mastering} leverages multi-modal large language models (MLLMs) to decompose complex prompts.

\subsection{Controllable Audio Generation}
Unlike the spatial control in image generation, recent efforts in controllable audio generation have primarily focused on enabling the precise specification of sequence, timestamp, and occurrence counts of events in the synthesized audio.
MC-Diffusion \cite{guo2024audio} stands out as an early attempt to introduce multi-scale conditioning, enabling simultaneous control over temporal and spectral properties of audio through dedicated encoders.
PicoAudio \cite{xie2025picoaudio} takes a different route by employing LLM to generate class-timestamp matrices, thereby achieving finer temporal alignment.
AudioComposer \cite{wang2025audiocomposer} adopts a more user-friendly, text-only interface for control, but this comes at event omissions and reduced temporal precision.
More recently, FreeAudio \cite{jiang2025freeaudio} introduces a training-free solution that aggregates attention maps at inference time, offering a new perspective on controllability, though it requires intricate parameter tuning and adds operational complexity.
These approaches collectively demonstrate an inherent trade-off triangle among three critical dimensions:
(1) precise temporal localization of events, (2) scalability to a large and diverse set of event classes, including open-vocabulary scenarios, and (3) the efficiency and simplicity of the overall system.
These fundamental limitations become the motivation for our work.

\subsection{Graph-based Representation Learning}
Graph neural networks \cite{scarselli2008graph, bronstein2021geometric, wu2020comprehensive} provide powerful frameworks for structured data modeling. 
\cite{kipf2017semi} introduced graph convolutional networks for neighbor aggregation, while \cite{velickovic2018graph} developed graph attention networks with adaptive weighting mechanisms.
Beyond GNNs, \cite{kim2022pure} demonstrated that pure transformer models can learn graph representations when equipped with structural encodings.
In visual scene understanding, graph representations model complex relationships effectively. \cite{xu2017scene} proposed iterative message passing for scene graph generation, and \cite{ji2020action} introduced Action Genome, representing actions as spatio-temporal scene graphs for dynamic scene understanding.
Capitalizing on its inherent strengths in modeling entities and their relational structure, we leverage graph-based representation learning for controllable audio generation in this paper.

\section{Method} \label{sec:method}
\begin{figure*}
\centering 
\includegraphics[width=1\textwidth]{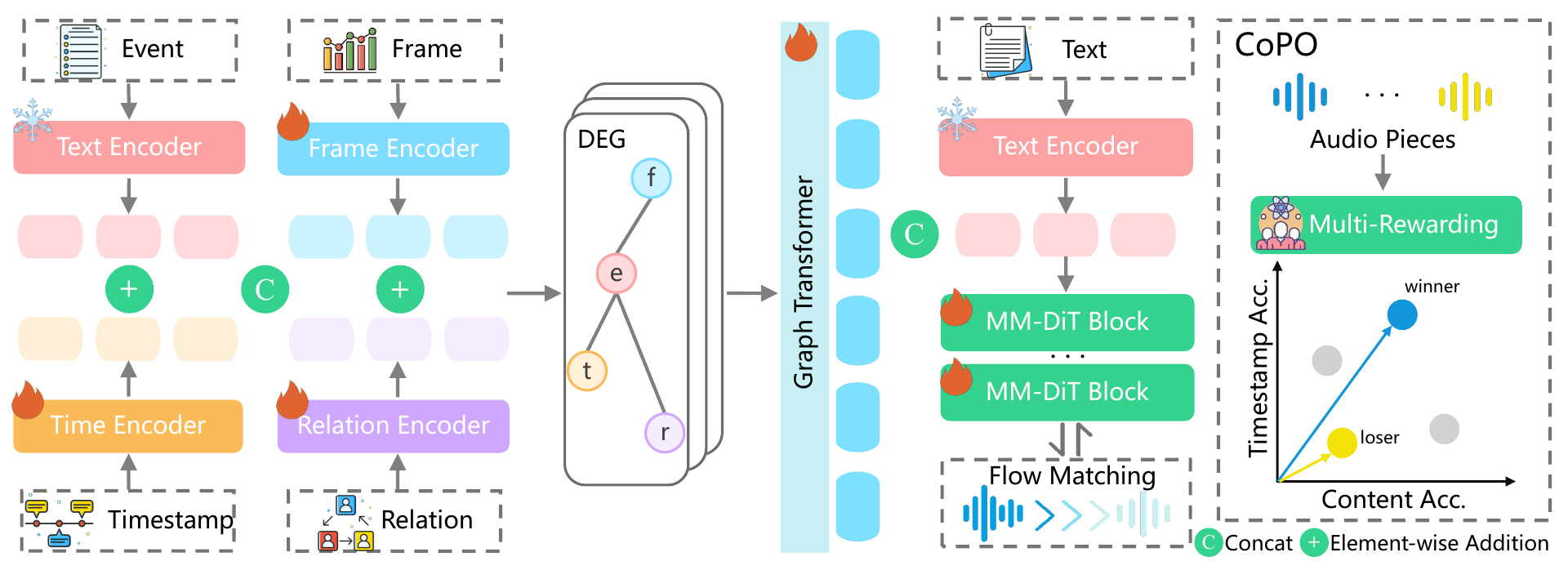}
\caption{The DegDiT overall framework decouples events from their timestamps and constructs frame-level event and relation matrices, which are subsequently mapped into structured graphs via dedicated encoders. A transformer aggregates these graphs to capture contextual dependencies, while a multi-modal diffusion transformer synthesizes output audio conditioned on both textual prompts and the structured graphs. The training pipeline is optimized with the flow matching loss followed by CoPO across multiple reward signals.} 
\label{Fig.main2} 
\end{figure*}

In this section, we introduce DegDiT, a novel framework for fine-grained, controllable text-to-audio generation based on the dynamic event graphs.
The goal of DegDiT is to synthesize audio that not only matches the semantic content of the text prompts but also adheres to detailed temporal specifications of sound events.
As shown in Fig. \ref{Fig.main2}, the overall framework constructs dynamic event graphs to represent audio events along with their temporal characteristics and relationships. 
These representations are then processed by a transformer to capture the contextual information.
Finally, we train a multi-modal diffusion transformer that generates audio conditioned on both text prompts and graph representations, ensuring semantic and temporal alignment with the input specifications.

\subsection{Dynamic Event Graph Representation}
Given a text prompt $T$, DegDiT constructs a dynamic event graph $\mathcal{G} = (\mathcal{V}, \mathcal{E})$, where:
\begin{itemize}
    \item $\mathcal{V}= \{v_1, v_2, \ldots, v_N\}$ is the set of all nodes representing audio events,
    \item $\mathcal{E} \subseteq \mathcal{V} \times \mathcal{V}$ is the set of edges representing interactions between events.
\end{itemize}
The node set $\mathcal{V}$ consists of $N$ audio events, where each $v_i \in \mathcal{V} $ is associated with a pair of specialized properties: an event property $\mathbf{E}_i$ and a temporal activation property $\mathbf{F}_i$.
The event property $\mathbf{E}_i$ serves as the anchor point, capturing the core attributes of each audio event through the tuple $(c_i, s_i, e_i, \alpha_i)$, where $c_i$ identifies the event category (such as ``dog barking''), while $s_i$ and $e_i$ mark its start and end times in seconds, and $\alpha_i \in [0,1]$ represents its intensity. 
This compact representation forms the basis for both semantic understanding and temporal localization of the events. 
On the other hand, the temporal activation property $\mathbf{F}$ tracks how the event manifests across time, which we formalize through a feature matrix $\mathbf{F} \in \mathbb{R}^{N \times F}$ that provides fine-grained control over event timing.
Here, each element $\mathbf{F}_{i,j}$ indicates how prominently event $i$ appears in time frame $j$, calculated by the proportion of duration within the frames.
Additionally, we define edges in $\mathcal{E}$ to capture temporal interactions between events, supporting five relationship types: \textit{before}, \textit{after}, \textit{overlaps}, \textit{contains}, and \textit{contained-by}.
These interactions are systematically captured in an adjacency tensor $\mathbf{A} \in \mathbb{R}^{N \times N \times 5}$, where each slice $\mathbf{A}_{i,j,:}$ measures the relationship between events $i$ and $j$.

To encode the textual description of the events, traditional approaches \cite{xie2025picoaudio} often rely on a fixed set of event types, limiting their ability to generalize to novel or rare events. 
In contrast, we employ an open-vocabulary text encoder based on a pretrained FLAN-T5 large language model \cite{chung2024scaling}, allowing flexible encoding of arbitrary textual event descriptions while leveraging the model's rich semantic knowledge. 
For each event, the category $c_i$ is firstly tokenized and passed through the text encoder:
\begin{equation}
\mathbf{h}^{\text{cat}}_i = \textrm{T5Encoder}(c_i) \in \mathbb{R}^{L \times d_{\text{T5}}},
\end{equation}
where $\mathbf{h}^{\text{cat}}_i$ is a sequence of token embeddings that capture the nuanced meaning of the event category, with $L$ being the sequence length and $d_{\text{T5}}$ the hidden dimension of the text encoder.
To obtain a fixed-length representation suitable for graph processing, we apply mean pooling along the sequence dimension, followed by a linear projection that adapts the embedding space to our specific task:
\begin{equation}
\mathbf{e}^{\text{cat}}_i = \textrm{Proj}(\textrm{MeanPool}(\mathbf{h}^{\text{cat}}_i)) \in \mathbb{R}^{d_{\text{emb}}}.
\end{equation}
This enables the model to capture the semantic information of the event categories while maintaining the flexibility to handle unseen event categories during inference.

After acquiring the semantic representation of each event, precise temporal modeling addresses the temporal characteristics of the sound event, specified by its onset and offset timestamps $(s_i, e_i)$. 
We implement this via a multi-layer perceptron (MLP) with sinusoidal activation functions that are particularly suited for representing periodic temporal patterns:
\begin{equation}
\mathbf{t}_i = \textrm{MLP}(\text{Concat}(s_i, e_i)) \in {\mathbb{R}^{d_{\text{emb}}}}.
\end{equation}
The final event representation $\mathbf{g}^{(0)}_i$ is obtained through element-wise addition of the semantic and temporal embeddings, combining all the aspects of event property attributes:
\begin{equation}
\mathbf{g}^{(0)}_i = \mathbf{e}^{\text{cat}}_i + \mathbf{t}_i \in \mathbb{R}^{d_{\text{emb}}}.
\end{equation}

In parallel, the temporal activity property for the pre-defined frames and each event is also encoded to capture the fine-grained temporal structure of audio events:
\begin{equation}
\mathbf{f}_i = \alpha_i \cdot \textrm{FrameEncoder}(\mathbf{F}_{i,:}) \in \mathbb{R}^{d_{\text{frame}}},
\end{equation}
where $\mathbf{F}_{i,:} \in \mathbb{R}^{F}$ is the temporal activation vector for event $i$ across all $F$ pre-defined frames, $\textrm{FrameEncoder}: \mathbb{R}^{F} \to \mathbb{R}^{d_{\text{frame}}}$ projects the frame-level features, and the intensity weighting $\alpha_i$ preserves the relative importance of different events. 
These features maintain precise alignment between the specified temporal structure and the generated audio.

Another key point in controllable audio generation lies in capturing the interplay between multiple sound events, which serves as the mechanism to achieve the coherence of audio events.
To capture these inter-event dependencies, we leverage the edge set $\mathcal{E}$ and its representation in the adjacency tensor $\mathbf{A}$.
Each event node $v_i$ aggregates contextual information from others through:
\begin{equation}
\mathbf{r}_i = \frac{1}{|\mathcal{N}_i|} \sum_{j \in \mathcal{N}_i} \textrm{RelationEncoder}(\mathbf{A}_{i,j,:}) \in \mathbb{R}^{d_{\text{rel}}},
\end{equation}
where $\mathcal{N}_i$ denotes the set of all other events, $\mathbf{A}_{i,j,:} \in \mathbb{R}^{5}$ encodes the edge features between $v_i$ and $v_j$, and $\textrm{RelationEncoder}: \mathbb{R}^{5} \to \mathbb{R}^{d_{\text{rel}}}$ is a learnable mapping. 
We set $d_{\text{rel}} = d_{\text{frame}}$ to ensure dimensional compatibility for subsequent fusion operations.

Finally, a learnable lightweight feed-forward network $\textrm{Fusion}(\cdot)$ integrates the event property features with relation and frame embeddings to produce the union representation:
\begin{equation}
\mathbf{g}^{(1)}_i = \textrm{Fusion}\big(\text{Concat}(\mathbf{g}^{(0)}_i, \mathbf{r}_i + \mathbf{f}_i)\big)\in {\mathbb{R}^{d_{\text{g}}}}.
\end{equation}
The collected representation of all the events  $\mathbf{G}^{(1)} = [\mathbf{g}^{(1)}_1; \mathbf{g}^{(1)}_2; \dots; \mathbf{g}^{(1)}_N] \in \mathbb{R}^{N \times d_g}$ is then processed by a transformer encoder, modeling both local and global dependencies among events through the self-attention mechanism:
\begin{equation}
\mathbf{H} = \textrm{TransformerEncoder}(\mathbf{G}^{(1)} + \textrm{PosEmb}, \mathbf{M}),
\end{equation}
where $\textrm{PosEmb} \in \mathbb{R}^{N \times d}$ adds positional information, and $\mathbf{M}$ is an optional attention mask.
The output $\mathbf{H}$ is a set of dynamic event graph embeddings that encode the structure and dynamics of the entire audio scene. 
Crucially, its embedding dimension aligns with the text embedding dimension of the diffusion model.

\begin{table*}[h]
\centering
\caption{Criterion, weights, and descriptions of the sample quality scoring function.}
\label{tab:quality_score}
\begin{tabular}{l l l }
\toprule
\textbf{Criterion} & \textbf{Weight} & \textbf{Description} \\
\midrule
Event count (2--5) & +10 & Ideal number of events per sample \\
Event count (1) & +5 & Single event \\
Event count (6--8) & +3 & Moderate number of events \\
Event count (others) & -5 & Too many or too few events \\
Type diversity ($\geq$ 3 classes) & +8 & Number of unique event classes \\
Type diversity (2 classes) & +5 & Moderate class diversity \\
Duration validity (0.5--5.0 s) & +2 per event (max 10) & Reasonable event duration \\
Ideal duration (1.0--3.0 s) & +1 per event (max 5) & Preferred event duration range \\
Short event penalty ($<$ min duration) & -50 & Severe penalty for very short events \\
Single event full coverage penalty & -8 & Penalty if single event covers $>$95 audio and is Speech or Music \\
\bottomrule
\end{tabular}
\end{table*}

\subsection{Training Strategy and Dataset Construction}
During training, we employ a diffusion transformer framework \cite{peebles2023scalable}, similar to TangoFlux \cite{hung2024tangoflux}, as the backbone of our model. 
To achieve both semantic and temporal alignment in audio synthesis, we condition the model on a concatenation of dynamic event graph embeddings $\mathbf{H}$, text prompt embeddings $\mathbf{E}_{\text{text}}$, and duration embeddings $\mathbf{E}_{\text{dur}}$:
\begin{equation}
\mathbf{E}_{\text{cond}} = \textrm{Concat}(\mathbf{E}_{\text{text}}, \mathbf{H}, \mathbf{E}_{\text{dur}}).
\end{equation}

For the training objective, we adopt the flow matching (FM) \cite{flowmatch}, which has demonstrated robustness to noise scheduling and improved sample efficiency in various text-to-audio tasks \cite{hung2024tangoflux,vyas2023audiobox}. 
Specifically, we utilize rectified flows \cite{liuflow}, where the forward process is defined as a straight-line interpolation between the clean audio latent $\mathbf{x}_1$ and a noise sample $\mathbf{x}_0 \sim \mathcal{N}(0, \mathbf{I})$ at a random timestep $t \in [0, 1]$:

\begin{equation}
\mathbf{z}_t = (1-t)\mathbf{x}_1 + t\mathbf{x}_0,
\end{equation}
with the corresponding target velocity given by:
\begin{equation}
\mathbf{u}_t = \frac{d\mathbf{z}_t}{dt} = \mathbf{x}_0 - \mathbf{x}_1.
\end{equation}
Enabling the model to learn a time-dependent vector field that transports samples from the noise distribution to the target audio distribution, we train the model $\epsilon_\theta(\mathbf{z}_t, t, \mathbf{E}_{\text{cond}})$ to predict this velocity using the flow matching loss:
\begin{equation}
\mathcal{L}_{\text{FM}} = \mathbb{E}_{\mathbf{x}_1, \mathbf{x}_0, t}\left[\|\epsilon_\theta(\mathbf{z}_t, t, \mathbf{E}_{\text{cond}}) - \mathbf{u}_t\|_2^2\right].
\end{equation}
The training procedure consists of a two-stage process: pre-training and subsequent fine-tuning, which jointly optimize the diffusion model and the training modules in DEG.
To prepare the pre-training dataset, we reprocess all AudioSet \cite{audioset} samples with HTS-AT \cite{chen2022hts}, a transformer-based model that achieves high-resolution event boundary extraction through token-semantic alignment. 
This process generates confidence-calibrated pseudo-labels for weakly annotated segments and refines temporal boundaries between audio tokens and label semantics.
However, training on this large dataset is time-consuming; thus, we design a Quality-Balanced Data Selection pipeline that integrates hierarchical audio event annotation with a multi-criteria quality optimization framework.
We randomly sample a 100k audio clip set $\mathcal{E}$ from the reprocessed dataset and analyze the distribution of event frequencies to determine event rarity thresholds adaptively. 
By default, an event is considered \emph{rare} if its relative frequency is below $0.5\%$, and \emph{common} if its relative frequency exceeds $3\%$. 
Formally, let \(N_\epsilon\) denote the total occurrences of event \(\epsilon\), and let \(\phi_\epsilon\) be its relative frequency:
\begin{equation}
\begin{aligned}
& \mathcal{E}_{\text{rare}} = \{ \epsilon \mid \phi_\epsilon < \tau_{\text{rare}} \}, \\
& \mathcal{E}_{\text{common}} = \{ \epsilon \mid \phi_\epsilon > \tau_{\text{common}} \}, \\
& \text{where} \quad \phi_\epsilon = \frac{N_\epsilon}{\sum_{\kappa \in \mathcal{E}} N_\kappa},
\end{aligned}
\end{equation}
These thresholds can also be adaptively computed from the data distribution to better reflect the characteristics of the dataset.
Based on the determined rarity thresholds, we assign each sample a quality score $Q(x)$ that reflects multiple criteria, including event count, type diversity, temporal precision, and event duration, as detailed in Tab. \ref{tab:quality_score}.

The final dataset construction adopts a stratified sampling approach that applies progressively stricter quality requirements as event frequency increases. Specifically, rare event included samples are retained with a moderate threshold ($Q \geq 10$) to preserve coverage; common event samples must satisfy the highest threshold ($Q \geq Q_{\min}$, $Q_{\min}=15$ by default) to maximize precision for dominant patterns; and medium-frequency ones are filtered with intermediate standards ($Q \geq 0.8\times Q_{\min}$) to balance diversity and reliability.
The curated dataset $\mathcal{D}'$ is partitioned into three mutually exclusive subsets based on event types and quality:
\begin{equation}
\begin{aligned}
&\mathcal{D}' = \mathcal{S}_{\text{rare}} \;\cup\; \mathcal{S}_{\text{common}} \;\cup\; \mathcal{S}_{\text{medium}}, \;\text{where} \\
&\left\{
\begin{array}{l}
\mathcal{S}_{\text{rare}} = \{ x \mid \exists x \in \mathcal{E}_{\text{rare}},\; Q(x) \geq 10 \}, \\
\mathcal{S}_{\text{common}} = \{ x \mid \exists x \in \mathcal{E}_{\text{common}},\; Q(x) \geq Q_{\min} \}, \\
\mathcal{S}_{\text{medium}} = \{ x \mid x \notin \mathcal{S}_{\text{rare}} \cup \mathcal{S}_{\text{common}},\; Q(x) \geq 0.8\,\times Q_{\min} \}
\end{array}
\right.
\end{aligned}
\label{eq:dataset_corrected}
\end{equation}

Let $N_{\mathcal{D}'}$ denotes the predefined sample numbers of $\mathcal{D}'$, with $0.25N_{\mathcal{D}'}$ samples selected from each of $\mathcal{S}_{\text{rare}}$ and $\mathcal{S}_{\text{medium}}$, and $0.5N_{\mathcal{D}'}$ samples from $\mathcal{S}_{\text{common}}$.
Afterwards, we pre-train the DegDiT model on the curated dataset $\mathcal{D}'$, enabling the model to learn robust representations from a large and diverse set of audio-text-event triplets. 
Subsequently, we fine-tune the model on the AudioCondition \cite{guo2024audio} training set, which provides high-quality event descriptions and temporal boundaries. 
This two-stage training procedure allows the model to benefit from both large-scale weakly labeled data and precise, expert-annotated audio examples.

\subsection{Consensus Preference Optimization}
Generative models are increasingly trained using reinforcement learning from human feedback (RLHF) \cite{dpo,azar2024general,lee2025calibrated} to better align model outputs with human preferences. 
In this paper, controllable audio generation poses unique challenges, as it requires balancing multiple reward criteria—including timestamp accuracy, event matching, and audio quality. 
To address this, we introduce a multi-reward reference optimization strategy designed to integrate these diverse objectives.
We first pick 5k audios and their corresponding prompts from AudioTime \cite{xie2025audiotime}, which is a manually generated dataset with accurate timestamps. 
Utilizing the same prompts, we generate 5k audios using TangoFlux \cite{hung2024tangoflux} and the fine-tuned DegDiT, respectively.
With these three parts of data, we extend the standard preference learning by incorporating fine-grained, multi-dimensional quality assessments named Consensus Preference Optimization (CoPO).
Unlike Direct Preference Optimization (DPO) \cite{dpo}, which treats preferences as binary signals, CoPO leverages continuous preference intensities derived from multiple reward metrics, since multiple evaluation criteria are available in controllable audio generation tasks.
In detail, the reward components are computed through specialized functions: text alignment via CLAP-score between audio and text embeddings ($r_{\text{text}}$), event alignment through comparison with parsed event descriptions ($r_{\text{event}}$), temporal accuracy against ground truth timestamps ($r_{\text{temporal}}$), and audio quality combining signal-to-noise ratio, spectral balance and dynamic range metrics ($r_{\text{audio}}$).  
These components are aggregated through the weights $\mathbf{w} = [w_{\text{text}}, w_{\text{event}}, w_{\text{temporal}}, w_{\text{audio}}]$ to form an overall quality score:
\begin{equation}
r_{\text{overall}}(x) = \mathbf{w}^T \mathbf{r}(x) = \sum_{i} w_i r_i(x).
\end{equation}

The preference intensity between samples is then defined as the quality difference $\delta = r_{\text{overall}}(x_{\text{win}}) - r_{\text{overall}}(x_{\text{lose}})$, capturing both the direction and magnitude of preference.
CoPO optimizes the model through a preference loss that aligns model predictions with target preference intensities. 
Given the current model $\pi_\theta$ and reference model $\pi_{\text{ref}}$, we compute the model's preference signal as:
\begin{equation}
\begin{split}
\hat{\delta}_\theta = 
    -\left(\mathcal{L}_{\text{FM}}^{\theta}(x_{\text{win}}) - \mathcal{L}_{\text{FM}}^{\theta}(x_{\text{lose}})\right) \\
    + \left(\mathcal{L}_{\text{FM}}^{\text{ref}}(x_{\text{win}}) - \mathcal{L}_{\text{FM}}^{\text{ref}}(x_{\text{lose}})\right).
\end{split}
\end{equation}
The complete CoPO objective combines preference learning loss with a flow matching objective:
\begin{equation}
\mathcal{L}_{\text{CoPO}} = \mathbb{E}_{(x_{\text{win}}, x_{\text{lose}}, p) \sim \mathcal{D}} \left[ (\delta - \beta \hat{\delta}_\theta)^2 \right] + \lambda \mathcal{L}_{\text{FM}}^{\theta}(x_{\text{win}}),
\end{equation}
with $\beta$ controlling sensitivity to preference differences and a trade-off parameter $\lambda$.
The training procedure iteratively alternates between two steps: computing multi-dimensional rewards and updating the model parameters. 
In parallel, the reference model is periodically refreshed through exponential moving averaging.

\begin{figure*}
\label{instruction}
\centering
\begin{minipage}{0.95\textwidth}
\begin{tcolorbox}[
    colback=blue!1!white,      
    colframe=blue!10!black,   
    coltitle=white,            
    colbacktitle=blue!1!black,
    fonttitle=\bfseries\large, 
    fontupper=\small,          
    boxrule=1pt,               
    arc=5pt,                   
    left=8pt, right=8pt, top=2pt, bottom=2pt,
    drop shadow,               
    enhanced,
    width=\textwidth,
]
Please listen to this \textbf{10.0}-second audio file and evaluate it based on the following requirements:

\textbf{EXPECTED CONTENT:}
\begin{itemize}[leftmargin=1em]
    \item \textbf{Overall Description:} ``A person says the phrase \emph{Hello, world!} clearly, followed by a short beep sound.''
    \item \textbf{Expected Temporal Segments:} 0.0s--2.5s: Speech,\ 6.0s--6.5s: Beep sound
\end{itemize}

\textbf{EVALUATION CRITERIA} (Rate each 1--10, where 10 is perfect):
\begin{enumerate}[leftmargin=1.5em, itemsep=0.3em]
    \item \textbf{CONTENT MATCHING:} Does the actual audio content match the expected descriptions?
    
     Are the right sounds/speech present? Does the content make sense with the description?
    \item \textbf{TIMING ACCURACY:} Do events occur at the correct times with precise boundaries?
    
     Are events happening when they should? Are start/end times accurate?
    \item \textbf{AUDIO QUALITY:} Is the audio technically good?
    
    Is it clear and audible? No artifacts, distortion, or quality issues?
    \item \textbf{OVERALL COHERENCE:} Does the entire audio work well together?

 Does it flow naturally? Does it match the overall expected description?
\end{enumerate}

\end{tcolorbox}
\end{minipage}

\caption{We provide Gemini 2.5 Pro with the evaluation template shown above, including overall descriptions and expected temporal segments (annotated from the ground truth), and request ratings across four dimensions: content matching, timing accuracy, audio quality, and overall coherence.}
\label{instruction} 
\end{figure*}

\section{Experiments}
\subsection{Implementations Details}
 \textbf{Datasets.}
We conduct our experiments using a diverse set of publicly available datasets to ensure comprehensive evaluation. 
For pre-training, we utilize AudioSet \cite{audioset}, a large-scale, weakly-labeled dataset comprising over 2 million 10-second YouTube video clips annotated with 527 sound event categories.
We apply the Quality-Balanced Data Selection pipeline to produce a curated subset of $N_{\mathcal{D}'} = 100,000$ high-quality samples for model pre-training.
For fine-tuning, we employ the AudioCondition training set \cite{guo2024audio}, including 89,557 strongly-labeled audio pieces with 439 event categories, which leverage audio and timestamp annotations from AudioSetStrong \cite{audiosetstrong}. 
During the CoPO phase, we leverage the AudioTime dataset \cite{xie2025audiotime}, which offers manually annotated audio samples with precise temporal information for sound events.
We select 5,000 samples with corresponding prompts and generate 5,000 synthetic audio samples using TangoFlux \cite{hung2024tangoflux} and our fine-tuned DegDiT model employing the same prompts to construct paired real and synthetic samples.
For evaluation, we utilize the test split of AudioCondition \cite{guo2024audio}, comprising 1,110 audio samples, as well as 692 samples from the DESED \cite{turpault2019desed} test set, which contains 10 sound event classes with strong temporal annotations for assessing temporal control accuracy. 
We also incorporate the test sets of AudioTime \cite{xie2025audiotime}, which offers high-quality audio with explicit event borders and a diverse range of event types, suitable for subjective evaluation.

\textbf{Evaluation Metrics.}
We evaluate the generated audio across three key dimensions: temporal accuracy, semantic content alignment, and perceptual audio quality, using both objective and subjective metrics. 
For temporal control, we report event-level (F1 Event) and clip-level (F1 Clip) scores based on the PB-SED \cite{ebbers2021forward,ebbers2021self,ebbers2022pre} sound event detection system.
F1 (Event) assesses frame-level precision and recall of event boundaries via segment-wise matching between generated and ground-truth events, while F1 (Clip) measures the presence or absence of target events at the clip level, macro-averaged across all classes.
Semantic alignment is evaluated using the Contrastive Language-Audio Pretraining (CLAP) model  \footnote{Checkpoint: music\_speech\_audioset\_epoch\_15\_esc\_89.98.pt, available at \url{https://github.com/LAION-AI/CLAP}.} \cite{clap}, reporting both CLAP (Audio)—the similarity between generated and reference audio in the embedding space—and CLAP (Text), which quantifies cosine similarity between audio embeddings and corresponding textual prompts, reflecting audio-text alignment. 
Audio quality is measured by PAM \cite{deshmukh2024pam}, a no-reference metric using audio-language models and text prompts to assess quality without reference data.
For AudioTime test set, we test the accuracy of duration, frequency, ordering, and timestamp, respectively, with its STEAMtool.

Since experiments in WoW-Bench \cite{kim2025wow} show that MLLMs perform comparably to human raters in judging audio categories and duration,
we design Audio Evaluation Instructions in Fig. \ref{instruction} and conduct subjective evaluations with both Gemini 2.5 Pro MLLM \cite{comanici2025gemini} and human raters. 
The generated samples are rated by the MLLM on content matching, timing accuracy, audio quality, and overall coherence. 
With the same four metrics, we also engage 17 domain experts in a subjective evaluation, where each expert independently rates 30 randomly selected samples from AudioTime \cite{xie2025audiotime} for each model.
The Mean Opinion Score (MOS) along with the 95\% confidence intervals are computed for the results.

\begin{table*}[htbp]
\centering
\caption{Experimental Results on \textbf{AudioCondition} Dataset (Objective Metrics and Gemini 2.5 Pro Rating). Boldface indicates the best performance, while underlining denotes the second-best.}
\begin{tabularx}{\textwidth}{l *{5}{>{\centering\arraybackslash}X} | *{4}{>{\centering\arraybackslash}X}}
\toprule

\multirow{2}{*}{Method} & \multicolumn{5}{c|}{\textbf{Objective Metrics}$\uparrow$} & \multicolumn{4}{c}{\textbf{Gemini 2.5 Pro Rating [score/10.00]}$\uparrow$} \\
\cmidrule(lr){2-6} \cmidrule(lr){7-10}
 & F1(Event) & F1(clip) & CLAP(A) & CLAP(T) & PAM & Content & Temporal & Quality & Overall \\
\midrule
Ground Truth & 0.409 & 0.655 & 1.000 & 0.156 & 0.672 & 7.82 & 8.95 & 8.48 & 8.16 \\
\hdashline
AudioLDM2 \cite{audioldm2-2024taslp} & 0.061 & 0.407 & 0.244 & \textbf{0.328} & 0.588 & 6.71 & 7.08 & 6.18 & 6.44 \\
Tango2 \cite{majumder2024tango} & 0.064 & 0.471 & 0.268 & \underline{0.325} & 0.231 & 4.16 & 4.63 & 4.78 & 4.02 \\
TangoFlux \cite{hung2024tangoflux} & 0.019 & 0.465 & 0.370 & 0.214 & 0.272 & 5.02 & 6.39 & 4.59 & 4.93 \\
\hdashline
TangoFlux (FT) \cite{hung2024tangoflux} & 0.423 & 0.675 & \underline{0.406} & 0.264 & \underline{0.822} & 8.63 & 9.31 & \underline{8.01} & 8.55 \\
TangoFlux+RP \cite{chen2024training} & 0.184 & 0.603 & 0.339 & 0.136 & 0.231 & 5.95 & 7.08 & 4.45 & 5.56 \\
MC-Diffusion \cite{guo2024audio} & 0.291 & 0.471 & - & - & - & - & - & - & - \\
AudioComposer \cite{wang2025audiocomposer} &\underline{0.466} & 0.608 & 0.236 & 0.238 & 0.641 & \textbf{9.22} & \textbf{9.76} & 7.11 & \underline{8.74} \\
FreeAudio \cite{jiang2025freeaudio} & 0.443 & \underline{0.685} & - & - & - & - & - & - & - \\
\rowcolor{lightblue} DegDiT(ours) & \textbf{0.589} & \textbf{0.846} & \textbf{0.428} & 0.270 & \textbf{0.830} & \underline{9.09} & \underline{9.50} & \textbf{8.56} & \textbf{9.01} \\
\bottomrule
\end{tabularx}
\label{tab:ac_results}
\end{table*}

\begin{table*}
\centering
\caption{Experimental Results on \textbf{DESED} Dataset (Objective Metrics and Gemini 2.5 Pro Rating). Boldface indicates the best performance, while underlining denotes the second-best.}
\begin{tabularx}{\textwidth}{l *{5}{>{\centering\arraybackslash}X} | *{4}{>{\centering\arraybackslash}X}}
\toprule
\multirow{2}{*}{Method} & \multicolumn{5}{c|}{\textbf{Objective Metrics}$\uparrow$} & \multicolumn{4}{c}{\textbf{Gemini 2.5 Pro Rating [score/10.00]}$\uparrow$} \\
\cmidrule(lr){2-6} \cmidrule(lr){7-10}
 & F1(Event) & F1(clip) & CLAP(A) & CLAP(T) & PAM & Content & Temporal & Quality & Overall \\
\midrule
Ground Truth & 0.666 & 0.882 & 1.000 & 0.201 & 0.646 & 9.62 & 9.74 & 9.57 & 9.67 \\
\hdashline
AudioLDM2 \cite{audioldm2-2024taslp} & 0.091 & 0.525 & 0.326 & \underline{0.271} & \underline{0.412} & 8.62 & 8.77 & 8.76 & 8.59 \\
Tango2 \cite{majumder2024tango} & 0.099 & 0.616 & 0.356 & 0.268 & 0.199 & 8.47 & 8.80 & 7.65 & 8.18 \\
TangoFlux \cite{hung2024tangoflux} & 0.094 & 0.621 & 0.422 & 0.213 & 0.248 & 7.38 & 8.31 & 8.14 & 7.57 \\
\hdashline
TangoFlux (FT) \cite{hung2024tangoflux} & \underline{0.378} & 0.564 & 0.413 & 0.222 & 0.403 & \underline{8.50} & \underline{9.19} & \underline{8.99} & \underline{8.73} \\
TangoFlux+RP \cite{chen2024training} & 0.170 & \underline{0.784} & \underline{0.516} & 0.192 & 0.297 & 8.48 & 9.07 & 8.62 & 8.56 \\
AudioComposer \cite{wang2025audiocomposer} & 0.339 & 0.519 & 0.308 & 0.197 & 0.242 & 7.86 & 9.09 & 8.92 & 8.33 \\
\rowcolor{lightblue} DegDiT(ours) & \textbf{0.612} & \textbf{0.784} & \textbf{0.520} & \textbf{0.279} & \textbf{0.490} & \textbf{9.08} & \textbf{9.45} & \textbf{9.15} & \textbf{9.15} \\
\bottomrule
\end{tabularx}
\label{tab:desed_results}
\end{table*}

\textbf{Implements.}
Several key hyperparameters are used throughout our experiments. The number of frames $F$ is set to 16, and the number of transformer layers $L$ is set to 4. 
In the CoPO module, the weighting vector $[w_{\text{text}}, w_{\text{event}}, w_{\text{temporal}}, w_{\text{audio}}]$ is set to $[0.35, 0.35, 0.15, 0.15]$.
Both the hyperparameters $\beta$ and $\lambda$ are set to 0.1.
To enable a comprehensive comparison, we reproduce various previous works as our baselines. 
We perform inference using the open-sourced pre-trained models AudioLDM2 \cite{audioldm2-2024taslp}, Tango2 \cite{majumder2024tango}, TangoFlux \cite{hung2024tangoflux} and AudioComposer \cite{wang2025audiocomposer}. 
TangoFlux \cite{hung2024tangoflux} is further fine-tuned on the AudioCondition dataset, referred to as TangoFlux (FT). 
The training-free method, Regional Prompting \cite{chen2024training}, is applied to TangoFlux \cite{hung2024tangoflux}, denoted as TangoFlux+RP. 
For non-open-sourced MC-Diffusion \cite{guo2024audio} and FreeAudio \cite{jiang2025freeaudio}, results from their original papers are reported.

During inference, we employ Classifier-Free Guidance (CFG) to enhance the semantic and temporal fidelity of the generated audio. The conditional model $\epsilon_\theta(\mathbf{z}_t, t, \mathbf{E}_{\text{cond}})$ and unconditional model $\epsilon_\theta(\mathbf{z}_t, t, \emptyset)$ are combined to compute the guided velocity prediction:
\begin{equation}
\begin{split}
\tilde{\epsilon}_\theta(\mathbf{z}_t, t, \mathbf{E}_{\text{cond}}) = 
    & \epsilon_\theta(\mathbf{z}_t, t, \emptyset) \\
    & + \gamma \cdot \left( \epsilon_\theta(\mathbf{z}_t, t, \mathbf{E}_{\text{cond}}) - \epsilon_\theta(\mathbf{z}_t, t, \emptyset) \right),
\end{split}
\end{equation}
where $\gamma$ denotes the guidance scale. 
We find $\gamma = 4.0$ to be optimal in our experiments. The inference process follows the rectified flow ODE, which is solved numerically using the guided velocity $\tilde{\epsilon}_\theta$ to produce audio samples.

\begin{table*}[h]
\centering
\caption{Comparison of Audio Generation Models on AudioTime Dataset and Evaluation metrics. Boldface indicates the best performance, while underlining denotes the second-best.}
\label{tab:audio_models}
\begin{tabular}{lcccc|cccc}
\toprule
\multirow{2}{*}{Method} & \multicolumn{4}{c|}{\textbf{Objective Metrics}} & \multicolumn{4}{c}{\textbf{Subjective Metrics [score/100.00]}$\uparrow$} \\
\cmidrule(lr){2-5} \cmidrule(lr){6-9}
 & Duration $\downarrow$ & Frequency $\downarrow$ & Ordering $\downarrow$ & TimeStamp $\uparrow$ & Content & Temporal & Quality & Overall \\
\midrule
Ground Truth & 0.793 & 0.342 & 0.224 & 0.906 & 97.57$\pm\scriptstyle0.39$ & 97.92$\pm\scriptstyle0.25$ & 96.32$\pm\scriptstyle0.46$ & 97.27$\pm\scriptstyle0.30$ \\
\midrule
AudioLDM2 \cite{audioldm2-2024taslp} & 3.404 & 1.639 & 0.960 & 0.543 & 71.41$\pm\scriptstyle3.69$ & 56.28$\pm\scriptstyle2.10$ & 70.20$\pm\scriptstyle2.00$ & 65.97$\pm\scriptstyle2.10$ \\
Tango2 \cite{majumder2024tango} & 3.695 & 1.521 & 0.858 & 0.609 & 68.23$\pm\scriptstyle1.97$ & 56.65$\pm\scriptstyle2.08$ & 68.26$\pm\scriptstyle1.81$ & 64.38$\pm\scriptstyle1.79$\\
TangoFlux \cite{hung2024tangoflux} & 3.302 &1.312 &  0.866& 0.613 & 70.79$\pm\scriptstyle1.88$ & 56.49$\pm\scriptstyle2.09$ & 68.81$\pm\scriptstyle1.87$ & 65.37$\pm\scriptstyle1.73$ \\
TangoFlux (FT) \cite{hung2024tangoflux} & 2.653 & 1.112 & \underline{0.766} & 0.665 & \underline{81.82$\pm\scriptstyle1.82$} & \underline{86.49$\pm\scriptstyle1.33$} & \underline{78.89$\pm\scriptstyle1.66$} & \underline{82.40$\pm\scriptstyle1.45$} \\
TangoFlux+RP \cite{chen2024training} & 3.509& 1.522 & 0.788 & \underline{0.671} & 78.28$\pm\scriptstyle1.56$ & 64.78$\pm\scriptstyle2.90$ & 76.68$\pm\scriptstyle1.54$ & 73.24$\pm\scriptstyle1.65$ \\
AudioComposer \cite{wang2025audiocomposer} & \textbf{2.166} & \underline{1.056} & 0.814 & 0.612 & 75.47$\pm\scriptstyle2.06$ & 84.44$\pm\scriptstyle1.74$ & 66.79$\pm\scriptstyle2.03$& 75.56$\pm\scriptstyle1.73$ \\
\rowcolor{lightblue} DegDiT(ours)& \underline{2.507} & \textbf{0.818} & \textbf{0.456} & \textbf{0.838} & \textbf{97.58$\pm\scriptstyle3.26$} & \textbf{96.01$\pm\scriptstyle0.44$} & \textbf{92.90$\pm\scriptstyle0.63$} & \textbf{95.50$\pm\scriptstyle1.14$} \\
\bottomrule
\end{tabular}
\end{table*}

\begin{figure*}
\centering 
\includegraphics[width=0.95\textwidth]{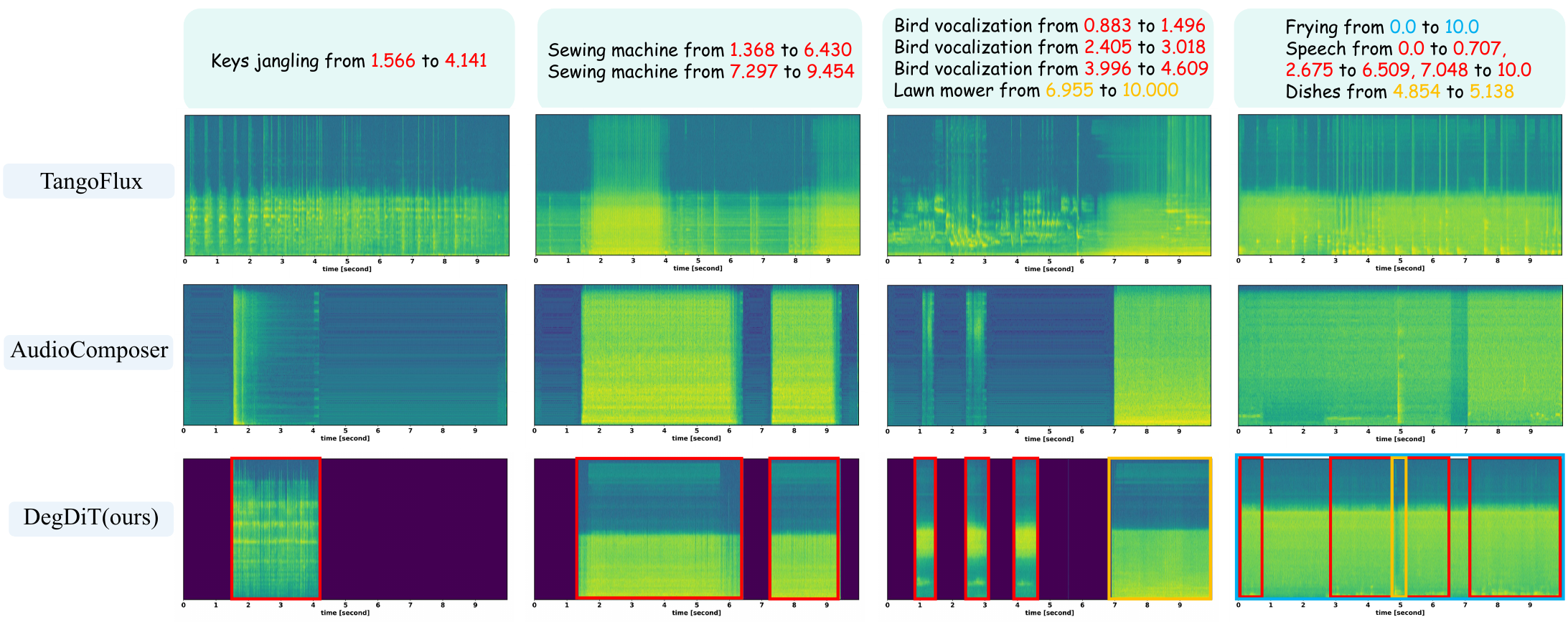}
\caption{We present examples of prompts containing single events, two events, multiple events, and overlapping events (in columns) for Tangoflux, Audiocomposer, and our proposed DegDiT (in rows), respectively.} 
\label{case} 
\end{figure*}

\subsection{Comparison with SOTA methods}
As shown in Tab. \ref{tab:ac_results} and \ref{tab:desed_results}, DegDiT exhibits consistent improvements over all baseline methods (sometimes even surpassing ground truth)  across a wide range of objective metrics on both AudioCondition and DESED datasets. 
Notably, DegDiT achieves the highest F1 (Event) and F1 (Clip) scores, highlighting the temporal precision in sound event detection. 
The model also attains superior CLAP (Audio) and CLAP (Text) scores, indicating a strong semantic correspondence between the generated outputs and reference signals. 
In addition, DegDiT leads in the audio quality metric PAM, confirming the naturalness and fidelity of the generated samples.

DegDiT’s performance also extends to evaluations by both large audio-language models and human evaluators.
According to the Gemini 2.5 Pro Rating, DegDiT consistently receives the highest or near-highest scores across content matching, timing
accuracy, and audio quality metrics on both datasets. 
Its elevated Overall scores suggest that the model not only excels in specific aspects but also provides a satisfying listening experience. 
By comparison, while methods such as AudioLDM2, TangoFlux (FT), and AudioComposer may perform competitively on certain individual metrics, they do not achieve the same level of balanced performance across all evaluation criteria. 
These findings underscore DegDiT’s effectiveness in advancing the state of controllable audio generation, both in terms of temporal accuracy and semantic alignment.

\begin{table}[t]
\centering
\scriptsize
\caption{Ablation study on AudioCondition Dataset}
\label{tab:ablation}
\begin{tabular}{lccccc}
\toprule
Module &  F1(Event) & F1(Clip) & CLAP(A) &CLAP(T) &PAM \\
\midrule
DiT (FT) & 0.423 & 0.675 & 0.406 & 0.264 & 0.822 \\
\hdashline
+DEG & 0.502 & 0.783 & 0.427 & 0.277 & 0.754 \\
+QBDS & 0.551 & 0.839 & 0.425 & 0.269 & \textbf{0.837}\\
+CoPO & \textbf{0.589} & \textbf{0.846} & \textbf{0.428} & 0.270 & 0.830 \\
\bottomrule
\end{tabular}
\end{table}

In Tab. \ref{tab:audio_models}, we present a comparison of the audio generation models on the AudioTime test set, evaluated across both objective and subjective metrics. 
The results demonstrate that DegDiT consistently achieves the best performance in the majority of objective metrics, including duration, frequency, ordering, and timestamp accuracy.
This indicates that DegDiT is highly effective at capturing the temporal structure and event alignment present in the reference audio.
Furthermore, DegDiT attains the highest scores in all subjective evaluation categories—content, temporal coherence, audio quality, and overall impression. 
Notably, the improvements in both objective and subjective metrics suggest that DegDiT produces audio that is natural and convincing to human listeners. 

In Fig. \ref{case}, we present representative examples of prompts containing single events, two events, multiple events, and overlapping events for three models: Tangoflux, Audiocomposer, and our proposed DegDiT.
Tangoflux does not incorporate explicit timestamp control signals, which limits its ability to generate audio with accurate temporal alignment of events. 
In contrast, Audiocomposer relies solely on textual descriptions for control. 
As a result, it sometimes fails to capture all events in multi-event scenarios and struggles to handle cases where events overlap in time. 
By comparison, our proposed DegDiT demonstrates robust performance across all these challenging conditions and generates audio that accurately reflects the specified event timing and content.

\subsection{Ablation Study}

\begin{table*}
\centering
\scriptsize
\caption{Performances under Different transformer layers ($L$) and pre-defined frame numbers ($F$). The row with a blue background is the default setting.}
\begin{tabular}{cc|ccccc|ccccc}
\toprule
\multicolumn{2}{c|}{Setting} & \multicolumn{5}{c|}{AudioCondition} & \multicolumn{5}{c}{Desed} \\
\cmidrule(r){1-2} \cmidrule(lr){3-7} \cmidrule(l){8-12}

Layers($L$) & Frames($F$) & F1(Event)$\uparrow$ & F1(Clip)$\uparrow$ & CLAP(A)$\uparrow$ & CLAP(T)$\uparrow$ & PAM$\uparrow$ & F1(Event)$\uparrow$ & F1(Clip)$\uparrow$ & CLAP(A)$\uparrow$ & CLAP(T)$\uparrow$ & PAM$\uparrow$ \\
\midrule
\rowcolor{lightblue}4 & 16 & 0.551 & 0.839 & 0.425 & 0.269 & 0.837 & 0.602 & 0.786 & 0.518 & 0.283 & 0.456 \\
\midrule
1 & 16 & 0.418 & 0.740 & 0.449 & 0.271 & 0.825 & 0.443 & 0.716 & 0.498 & 0.272 & 0.493 \\
2 & 16 & 0.591 & 0.791 & 0.421 & 0.261 & 0.833 & 0.618 & 0.793 & 0.504 & 0.279 & 0.500 \\
8 & 16 & 0.543 & 0.804 & 0.431 & 0.260 & 0.841 & 0.561 & 0.768 & 0.500 & 0.275 & 0.483 \\
16 & 16 & 0.550 & 0.750 & 0.431 & 0.265 & 0.844 & 0.521 & 0.718 & 0.485 & 0.267 & 0.464 \\
\midrule
4 & 32 & 0.523 & 0.782 & 0.423 & 0.275 & 0.816 & 0.552 & 0.745 & 0.498 & 0.275 & 0.443 \\
4 & 64 & 0.548 & 0.743 & 0.425 & 0.259 & 0.820 & 0.584 & 0.744 & 0.506 & 0.277 & 0.476 \\
4 & 128 & 0.519 & 0.755 & 0.423 & 0.272 & 0.830 & 0.523 & 0.730 & 0.490 & 0.275 & 0.468\\
4 & 256 & 0.519 & 0.791 & 0.440 & 0.273 & 0.817 & 0.559 & 0.766 & 0.506 & 0.269 & 0.440 \\
\bottomrule
\end{tabular}
\label{tab:layer}
\end{table*}

\begin{table*}[h]
\centering
\scriptsize
\caption{Performances under Different Guidance Scale (GS) and denoising Step Settings during inference. The row with a blue background is the default setting.}
\begin{tabular}{cc|ccccc|ccccc}
\toprule
\multicolumn{2}{c|}{Setting} & \multicolumn{5}{c|}{AudioCondition} & \multicolumn{5}{c}{Desed} \\
\cmidrule(r){1-2} \cmidrule(lr){3-7} \cmidrule(l){8-12}
$\gamma$ & Steps & F1(Event)$\uparrow$ & F1(Clip)$\uparrow$ & CLAP(A)$\uparrow$ & CLAP(T)$\uparrow$ & PAM$\uparrow$ & F1(Event)$\uparrow$ & F1(Clip)$\uparrow$ & CLAP(A)$\uparrow$ & CLAP(T)$\uparrow$ & PAM$\uparrow$ \\
\midrule
\rowcolor{lightblue}4 & 50 & 0.551 & 0.839 & 0.425 & 0.269 & 0.837 & 0.602 & 0.786 & 0.518 & 0.283 & 0.456 \\
\midrule
0 & 50 & 0.299 & 0.518 & 0.367 & 0.261 & 0.682 & 0.320 & 0.569 & 0.393 & 0.207 & 0.294 \\
2 & 50 & 0.498 & 0.749 & 0.425 & 0.279 & 0.800 & 0.526 & 0.710 & 0.488 & 0.257 & 0.428 \\
6 & 50 & 0.581 & 0.806 & 0.420 & 0.263 & 0.819 & 0.611 & 0.774 & 0.520 & 0.275 & 0.503 \\
8 & 50 & 0.565 & 0.792 & 0.408 & 0.255 & 0.796 & 0.599 & 0.764 & 0.507 & 0.271 & 0.495 \\
\midrule
4 & 10 & 0.497 & 0.754 & 0.404 & 0.279 & 0.800 & 0.575 & 0.741 & 0.450 & 0.253 & 0.409 \\
4 & 25 & 0.566 & 0.826 & 0.426 & 0.272 & 0.823 & 0.610 & 0.777 & 0.511 & 0.276 & 0.478 \\
4 & 100 & 0.598 & 0.857 & 0.429 & 0.268 & 0.829 & 0.610 & 0.787 & 0.526 & 0.277 & 0.493 \\
4 & 200 & 0.589 & 0.842 & 0.429 & 0.267 & 0.829 & 0.611 & 0.786 & 0.527 & 0.276 & 0.497 \\
\bottomrule
\end{tabular}
\label{tab:Guidance}
\end{table*}

In this section, we present the ablation study and parameter sensitivity experiments for our proposed method. 
As shown in Tab.~\ref{tab:ablation}, the progressive increases in both F1 (Event) and F1 (Clip) scores demonstrate the effectiveness of each proposed module.
We initially train the diffusion transformer on the AudioCondition dataset, denoted as DiT (FT), following the fine-tuning procedure described earlier.
The substantial improvement following the introduction of Dynamic Event Graphs (DEG) underscores the importance of modeling temporal dependencies and event interactions, enabling the system to more accurately capture the complex structure of audio events. 
Further gains achieved through the Quality-Balanced Data Selection (QBDS) pipeline suggest that careful curation of training data can mitigate the adverse effects of noise and class imbalance. 
Finally, the additional improvements observed with Consensus Preference Optimization (CoPO) indicate that leveraging the consensus among multiple reward models leads to comprehensive enhancements in model performance. 
Consequently, our findings clarify that the DEG embedding cannot function as a standalone control signal, as experiments show that training without a text prompt leads to non-convergence.

During training, the number of graph transformer layers $L$ and the frame $F$ are the key architectural and data-related hyperparameters that might influence the model’s capacity. 
As presented in Tab. \ref{tab:layer}, variations in either $L$ or $F$ do not lead to consistent improvements or declines across evaluation metrics. 
The results demonstrate the model's stable performance in modeling event correlations and temporal dependencies irrespective of hyperparameter variations.
Such robustness ensures consistent performance across varying configurations, enabling adaptable deployment in diverse scenarios and resource constraints without sacrificing performance.

When performing inference with diffusion models, adjusting the guidance scale and the number of diffusion steps can influence output fidelity, diversity, and computational resources required.
According to Tab. \ref{tab:Guidance}, when the guidance scale $\gamma$ is set to zero, the model fails to produce the target audio under DEG guidance. 
However, as $\gamma$ increases within a reasonable range, only moderate changes are observed in event-level and clip-level F1 scores, as well as in perceptual and audio-text alignment metrics. 
Similarly, varying the number of diffusion denoising steps does not lead to significant changes in any of the reported metrics. 
Across both the AudioCondition and DESED datasets, the model maintains robust performance under all tested configurations, with only slight variations. 
These results indicate that the model’s effectiveness in generating high-quality, semantically accurate audio remains stable, even when inference-time parameters are adjusted within a reasonable range.

\begin{table}
\centering
\caption{Comparison of Model Complexity and Inference Efficiency.}
\begin{tabular}{lccc}
\toprule
Method & Params (M) & GPU Mem (GB) & Time (s)  \\
\midrule
AudioLDM2 & 346 & 2.46 & 17.54 \\
Tango2 & 866 & 5.39 & 17.35 \\
TangoFlux* & 515 & 5.11 & 2.76 \\
AudioComposer & 272 & 10.22 & 1.56 \\
\rowcolor{lightblue} DegDiT (Ours) & 568 & 5.31 & 2.87 \\
\bottomrule
\end{tabular}
\label{tab:efficiency_comparison}
\end{table}

\subsection{Computational Requirements}
In this part, we compare model complexity and inference efficiency against the baseline methods in Tab. \ref{tab:efficiency_comparison}. 
The evaluation was conducted on 100 prompts from the AudioTime dataset using a single V100 GPU, with GPU memory representing peak usage and time denoting average latency. 
Notably, the computational profiles of TangoFlux+RP and TangoFlux (FT) closely resemble the base TangoFlux model, hence we employ TangoFlux* as their representative. Our DegDiT is suitable for deployment on consumer-grade 12GB GPUs. 
For training efficiency, we used an 8$\times$V100 cluster with a batch size of 6 per GPU. 
The pre-training stage on our 100k quality-balanced dataset required approximately 3 days, followed by 2 days of fine-tuning on the AudioCondition \cite{guo2024audio} Dataset.
Training with CoPO typically converges in just a few epochs, which takes less than 1 hour.

\section{Conclusion and future work}
In conclusion, we propose DegDiT, a dynamic event graph guided diffusion transformer framework for controllable text-to-audio generation. 
By encoding multi-event prompts as structured event graphs and modeling their temporal relationships, DegDiT enables fine-grained control over generated audio content. 
The introduction of a quality-balanced data selection pipeline and consensus preference optimization further enhances data diversity and multi-reward alignment.
Experimental results on several benchmarks demonstrate that DegDiT achieves state-of-the-art performance in terms of temporal precision and usability.

Despite its strong performance, DegDiT occasionally generates redundant audio segments when faced with scenarios involving the minority event classes \cite{chen2023area}. 
To understand the causes, we conducted an analysis based on Stable Flow \cite{avrahami2025stable}. It reveals that while generating these minority events, attention leakage occurs in later steps and deeper transformer layers, leading to redundant output.
Due to this, our future work will focus on constructing a large-scale dataset with precise timestamp annotations covering a diverse array of events. 
We anticipate that such a dataset will not only enhance the performance of controllable text-to-audio generation, especially for rare events, but also facilitate more robust and reliable audio synthesis across a broader range of real-world scenarios.

\bibliographystyle{IEEEtran}
\bibliography{cas-refs} 

@inproceedings{chen2022hts,
  title={HTS-AT: A hierarchical token-semantic audio transformer for sound classification and detection},
  author={Chen, Ke and Du, Xingjian and Zhu, Bilei and Ma, Zejun and Berg-Kirkpatrick, Taylor and Dubnov, Shlomo},
  booktitle={ICASSP 2022-2022 IEEE International Conference on Acoustics, Speech and Signal Processing (ICASSP)},
  pages={646--650},
  year={2022},
  organization={IEEE}
}

@article{audioldm2-2024taslp,
  author={Liu, Haohe and Yuan, Yi and Liu, Xubo and Mei, Xinhao and Kong, Qiuqiang and Tian, Qiao and Wang, Yuping and Wang, Wenwu and Wang, Yuxuan and Plumbley, Mark D.},
  journal={IEEE/ACM Transactions on Audio, Speech, and Language Processing}, 
  title={AudioLDM 2: Learning Holistic Audio Generation With Self-Supervised Pretraining}, 
  year={2024},
  volume={32},
  pages={2871-2883},
  doi={10.1109/TASLP.2024.3399607}
}

@inproceedings{majumder2024tango,
  title={Tango 2: Aligning diffusion-based text-to-audio generations through direct preference optimization},
  author={Majumder, Navonil and Hung, Chia-Yu and Ghosal, Deepanway and Hsu, Wei-Ning and Mihalcea, Rada and Poria, Soujanya},
  booktitle={Proceedings of the 32nd ACM International Conference on Multimedia},
  pages={564--572},
  year={2024}
}

@inproceedings{guo2024audio,
  title={Audio generation with multiple conditional diffusion model},
  author={Guo, Zhifang and Mao, Jianguo and Tao, Rui and Yan, Long and Ouchi, Kazushige and Liu, Hong and Wang, Xiangdong},
  booktitle={Proceedings of the AAAI Conference on Artificial Intelligence},
  volume={38},
  number={16},
  pages={18153--18161},
  year={2024}
}

@inproceedings{xie2025picoaudio,
  title={PicoAudio: Enabling Precise Temporal Controllability in Text-to-Audio Generation},
  author={Xie, Zeyu and Xu, Xuenan and Wu, Zhizheng and Wu, Mengyue},
  booktitle={ICASSP 2025-2025 IEEE International Conference on Acoustics, Speech and Signal Processing (ICASSP)},
  pages={1--5},
  year={2025},
  organization={IEEE}
}

@inproceedings{wang2025audiocomposer,
  title={AudioComposer: Towards Fine-grained Audio Generation with Natural Language Descriptions},
  author={Wang, Yuanyuan and Chen, Hangting and Yang, Dongchao and Wu, Zhiyong and Wu, Xixin},
  booktitle={ICASSP 2025-2025 IEEE International Conference on Acoustics, Speech and Signal Processing (ICASSP)},
  pages={1--5},
  year={2025},
  organization={IEEE}
}

@article{hung2024tangoflux,
  title={Tangoflux: Super fast and faithful text to audio generation with flow matching and clap-ranked preference optimization},
  author={Hung, Chia-Yu and Majumder, Navonil and Kong, Zhifeng and Mehrish, Ambuj and Bagherzadeh, Amir Ali and Li, Chuan and Valle, Rafael and Catanzaro, Bryan and Poria, Soujanya},
  journal={arXiv preprint arXiv:2412.21037},
  year={2024}
}

@article{chen2024training,
  title={Training-free Regional Prompting for Diffusion Transformers},
  author={Chen, Anthony and Xu, Jianjin and Zheng, Wenzhao and Dai, Gaole and Wang, Yida and Zhang, Renrui and Wang, Haofan and Zhang, Shanghang},
  journal={arXiv preprint arXiv:2411.02395},
  year={2024}
}

@inproceedings{jiang2025freeaudio,
  title     = {FreeAudio: Training-Free Timing Planning for Controllable Long-Form Text-to-Audio Generation},
  author    = {Yuxuan Jiang and Zehua Chen and Zeqian Ju and Chang Li and Weibei Dou and Jun Zhu},
  booktitle = {Proceedings of the 2025 ACM International Conference on Multimedia (ACM MM)},
  year      = {2025}
}

@article{croitoru2023diffusion,
  title={Diffusion models in vision: A survey},
  author={Croitoru, Florinel-Alin and Hondru, Vlad and Ionescu, Radu Tudor and Shah, Mubarak},
  journal={IEEE transactions on pattern analysis and machine intelligence},
  volume={45},
  number={9},
  pages={10850--10869},
  year={2023},
  publisher={Ieee}
}

@inproceedings{rombach2022high,
  title={High-resolution image synthesis with latent diffusion models},
  author={Rombach, Robin and Blattmann, Andreas and Lorenz, Dominik and Esser, Patrick and Ommer, Bj{\"o}rn},
  booktitle={Proceedings of the IEEE/CVF conference on computer vision and pattern recognition},
  pages={10684--10695},
  year={2022}
}

@inproceedings{songscore,
  title={Score-Based Generative Modeling through Stochastic Differential Equations},
  author={Song, Yang and Sohl-Dickstein, Jascha and Kingma, Diederik P and Kumar, Abhishek and Ermon, Stefano and Poole, Ben},
  booktitle={International Conference on Learning Representations},
year={2021}
}

@misc{stabilityai,
  author       = {StabilityAI},
  title        = {Stable Diffusion},
  howpublished = {Hugging Face Model Hub},
  year         = {2024},
  url          = {https://huggingface.co/stabilityai},
}

@misc{flux,
  author       = {Black-Forest-Labs},
  title        = {FLUX.1-dev},
  howpublished = {Hugging Face Model Hub},
  year         = {2024},
  url          = {https://huggingface.co/black-forest-labs/FLUX.1-dev},
}

@misc{sora2024,
  author = {OpenAI},
  title = {Sora: Creating Video from Text},
  year = {2024},
  url          =  {https://openai.com/sora},
}

@inproceedings{karras2020analyzing,
  title={Analyzing and improving the image quality of stylegan},
  author={Karras, Tero and Laine, Samuli and Aittala, Miika and Hellsten, Janne and Lehtinen, Jaakko and Aila, Timo},
  booktitle={Proceedings of the IEEE/CVF conference on computer vision and pattern recognition},
  pages={8110--8119},
  year={2020}
}

@article{xiong2024autoregressive,
  title={Autoregressive models in vision: A survey},
  author={Xiong, Jing and Liu, Gongye and Huang, Lun and Wu, Chengyue and Wu, Taiqiang and Mu, Yao and Yao, Yuan and Shen, Hui and Wan, Zhongwei and Huang, Jinfa and others},
  journal={arXiv preprint arXiv:2411.05902},
  year={2024}
}

@article{goodfellow2014generative,
  title={Generative adversarial nets},
  author={Goodfellow, Ian J and Pouget-Abadie, Jean and Mirza, Mehdi and Xu, Bing and Warde-Farley, David and Ozair, Sherjil and Courville, Aaron and Bengio, Yoshua},
  journal={Advances in neural information processing systems},
  volume={27},
  year={2014}
}

@article{liu2023audioldm,
  title={Audioldm: Text-to-audio generation with latent diffusion models},
  author={Liu, Haohe and Chen, Zehua and Yuan, Yi and Mei, Xinhao and Liu, Xubo and Mandic, Danilo and Wang, Wenwu and Plumbley, Mark D},
  journal={arXiv preprint arXiv:2301.12503},
  year={2023}
}

@inproceedings{huang2023make,
  title={Make-an-audio: Text-to-audio generation with prompt-enhanced diffusion models},
  author={Huang, Rongjie and Huang, Jiawei and Yang, Dongchao and Ren, Yi and Liu, Luping and Li, Mingze and Ye, Zhenhui and Liu, Jinglin and Yin, Xiang and Zhao, Zhou},
  booktitle={International Conference on Machine Learning},
  pages={13916--13932},
  year={2023},
  organization={PMLR}
}

@inproceedings{zhang2023adding,
  title={Adding conditional control to text-to-image diffusion models},
  author={Zhang, Lvmin and Rao, Anyi and Agrawala, Maneesh},
  booktitle={Proceedings of the IEEE/CVF international conference on computer vision},
  pages={3836--3847},
  year={2023}
}

@inproceedings{yuan2025identity,
  title={Identity-preserving text-to-video generation by frequency decomposition},
  author={Yuan, Shenghai and Huang, Jinfa and He, Xianyi and Ge, Yunyang and Shi, Yujun and Chen, Liuhan and Luo, Jiebo and Yuan, Li},
  booktitle={Proceedings of the Computer Vision and Pattern Recognition Conference},
  pages={12978--12988},
  year={2025}
}

@article{guo2024pulid,
  title={Pulid: Pure and lightning id customization via contrastive alignment},
  author={Guo, Zinan and Wu, Yanze and Zhuowei, Chen and Zhang, Peng and He, Qian and others},
  journal={Advances in neural information processing systems},
  volume={37},
  pages={36777--36804},
  year={2024}
}

@article{kreuk2022audiogen,
  title={Audiogen: Textually guided audio generation},
  author={Kreuk, Felix and Synnaeve, Gabriel and Polyak, Adam and Singer, Uriel and D{\'e}fossez, Alexandre and Copet, Jade and Parikh, Devi and Taigman, Yaniv and Adi, Yossi},
  journal={arXiv preprint arXiv:2209.15352},
  year={2022}
}

@inproceedings{peebles2023scalable,
  title={Scalable diffusion models with transformers},
  author={Peebles, William and Xie, Saining},
  booktitle={Proceedings of the IEEE/CVF international conference on computer vision},
  pages={4195--4205},
  year={2023}
}

@inproceedings{ghosal2023text,
  title={Text-to-audio generation using instruction guided latent diffusion model},
  author={Ghosal, Deepanway and Majumder, Navonil and Mehrish, Ambuj and Poria, Soujanya},
  booktitle={Proceedings of the 31st ACM International Conference on Multimedia},
  pages={3590--3598},
  year={2023}
}

@article{xue2024auffusion,
  title={Auffusion: Leveraging the power of diffusion and large language models for text-to-audio generation},
  author={Xue, Jinlong and Deng, Yayue and Gao, Yingming and Li, Ya},
  journal={IEEE/ACM Transactions on Audio, Speech, and Language Processing},
  year={2024},
  publisher={IEEE}
}

@inproceedings{audioset,
  title={Audio set: An ontology and human-labeled dataset for audio events},
  author={Gemmeke, Jort F and Ellis, Daniel PW and Freedman, Dylan and Jansen, Aren and Lawrence, Wade and Moore, R Channing and Plakal, Manoj and Ritter, Marvin},
  booktitle={2017 IEEE international conference on acoustics, speech and signal processing (ICASSP)},
  pages={776--780},
  year={2017},
  organization={IEEE}
}

@dataset{turpault2019desed,
  author = {Turpault, Nicolas and Serizel, Romain and Shah, Ankit and Salamon, Justin},
  title = {{Sound Event Detection in Domestic Environments (DESED\_public\_eval)}},
  year = {2019},
  month = dec,
  publisher = {Zenodo},
  doi = {10.5281/zenodo.3588172},
  url = {https://doi.org/10.5281/zenodo.3588172},
}

@inproceedings{xie2025audiotime,
  title={Audiotime: A temporally-aligned audio-text benchmark dataset},
  author={Xie, Zeyu and Xu, Xuenan and Wu, Zhizheng and Wu, Mengyue},
  booktitle={ICASSP 2025-2025 IEEE International Conference on Acoustics, Speech and Signal Processing (ICASSP)},
  pages={1--5},
  year={2025},
  organization={IEEE}
}

@article{chung2024scaling,
  title={Scaling instruction-finetuned language models},
  author={Chung, Hyung Won and Hou, Le and Longpre, Shayne and Zoph, Barret and Tay, Yi and Fedus, William and Li, Yunxuan and Wang, Xuezhi and Dehghani, Mostafa and Brahma, Siddhartha and others},
  journal={Journal of Machine Learning Research},
  volume={25},
  number={70},
  pages={1--53},
  year={2024}
}

@article{ye2023ip,
  title={Ip-adapter: Text compatible image prompt adapter for text-to-image diffusion models},
  author={Ye, Hu and Zhang, Jun and Liu, Sibo and Han, Xiao and Yang, Wei},
  journal={arXiv preprint arXiv:2308.06721},
  year={2023}
}

@inproceedings{yang2024mastering,
  title={Mastering text-to-image diffusion: Recaptioning, planning, and generating with multimodal llms},
  author={Yang, Ling and Yu, Zhaochen and Meng, Chenlin and Xu, Minkai and Ermon, Stefano and Cui, Bin},
  booktitle={Forty-first International Conference on Machine Learning},
  year={2024}
}

@article{dpo,
  title={Direct preference optimization: Your language model is secretly a reward model},
  author={Rafailov, Rafael and Sharma, Archit and Mitchell, Eric and Manning, Christopher D and Ermon, Stefano and Finn, Chelsea},
  journal={Advances in neural information processing systems},
  volume={36},
  pages={53728--53741},
  year={2023}
}

@inproceedings{audiosetstrong,
  title={The benefit of temporally-strong labels in audio event classification},
  author={Hershey, Shawn and Ellis, Daniel PW and Fonseca, Eduardo and Jansen, Aren and Liu, Caroline and Moore, R Channing and Plakal, Manoj},
  booktitle={ICASSP 2021-2021 IEEE International Conference on Acoustics, Speech and Signal Processing (ICASSP)},
  pages={366--370},
  year={2021},
  organization={IEEE}
}

@inproceedings{clap,
  title={Large-scale contrastive language-audio pretraining with feature fusion and keyword-to-caption augmentation},
  author={Wu, Yusong and Chen, Ke and Zhang, Tianyu and Hui, Yuchen and Berg-Kirkpatrick, Taylor and Dubnov, Shlomo},
  booktitle={ICASSP 2023-2023 IEEE International Conference on Acoustics, Speech and Signal Processing (ICASSP)},
  pages={1--5},
  year={2023},
  organization={IEEE}
}

@article{ebbers2021forward,
  title={Forward-backward convolutional recurrent neural networks and tag-conditioned convolutional neural networks for weakly labeled semi-supervised sound event detection},
  author={Ebbers, Janek and Haeb-Umbach, Reinhold},
  journal={arXiv preprint arXiv:2103.06581},
  year={2021}
}

@inproceedings{ebbers2021self,
  title={Self-trained audio tagging and sound event detection in domestic environments},
  author={Ebbers, Janek and Haeb-Umbach, Reinhold},
  booktitle={Proceedings of the 6th Detection and Classification of Acoustic Scenes and Events 2021 Workshop (DCASE2021)},
  year={2021}
}

@article{ebbers2022pre,
  title={Pre-training and self-training for sound event detection in domestic environments},
  author={Ebbers, Janek and Haeb-Umbach, Reinhold},
  journal={arXiv preprint},
  year={2022}
}

@inproceedings{deshmukh2024pam,
  title={PAM: Prompting Audio-Language Models for Audio Quality Assessment},
  author={Deshmukh, Soham and Alharthi, Dareen and Elizalde, Benjamin and Gamper, Hannes and Al Ismail, Mahmoud and Singh, Rita and Raj, Bhiksha and Wang, Huaming},
  booktitle={Proc. Interspeech 2024},
  pages={3320--3324},
  year={2024}
}

@article{comanici2025gemini,
  title={Gemini 2.5: Pushing the frontier with advanced reasoning, multimodality, long context, and next generation agentic capabilities},
  author={Comanici, Gheorghe and Bieber, Eric and Schaekermann, Mike and Pasupat, Ice and Sachdeva, Noveen and Dhillon, Inderjit and Blistein, Marcel and Ram, Ori and Zhang, Dan and Rosen, Evan and others},
  journal={arXiv preprint arXiv:2507.06261},
  year={2025}
}

@inproceedings{flowmatch,
  title={Flow Matching for Generative Modeling},
  author={Lipman, Yaron and Chen, Ricky TQ and Ben-Hamu, Heli and Nickel, Maximilian and Le, Matt},
  booktitle={11th International Conference on Learning Representations, ICLR 2023},
  year={2023}
}

@inproceedings{liuflow,
  title={Flow Straight and Fast: Learning to Generate and Transfer Data with Rectified Flow},
  author={Liu, Xingchao and Gong, Chengyue and others},
  booktitle={The Eleventh International Conference on Learning Representations},
year={2023}
}

@article{vyas2023audiobox,
  title={Audiobox: Unified audio generation with natural language prompts},
  author={Vyas, Apoorv and Shi, Bowen and Le, Matthew and Tjandra, Andros and Wu, Yi-Chiao and Guo, Baishan and Zhang, Jiemin and Zhang, Xinyue and Adkins, Robert and Ngan, William and others},
  journal={arXiv preprint arXiv:2312.15821},
  year={2023}
}

@inproceedings{lee2025calibrated,
  title={Calibrated multi-preference optimization for aligning diffusion models},
  author={Lee, Kyungmin and Li, Xiahong and Wang, Qifei and He, Junfeng and Ke, Junjie and Yang, Ming-Hsuan and Essa, Irfan and Shin, Jinwoo and Yang, Feng and Li, Yinxiao},
  booktitle={Proceedings of the Computer Vision and Pattern Recognition Conference},
  pages={18465--18475},
  year={2025}
}

@article{hai2024ezaudio,
  title={Ezaudio: Enhancing text-to-audio generation with efficient diffusion transformer},
  author={Hai, Jiarui and Xu, Yong and Zhang, Hao and Li, Chenxing and Wang, Helin and Elhilali, Mounya and Yu, Dong},
  journal={arXiv preprint arXiv:2409.10819},
  year={2024}
}

@inproceedings{azar2024general,
  title={A general theoretical paradigm to understand learning from human preferences},
  author={Azar, Mohammad Gheshlaghi and Guo, Zhaohan Daniel and Piot, Bilal and Munos, Remi and Rowland, Mark and Valko, Michal and Calandriello, Daniele},
  booktitle={International Conference on Artificial Intelligence and Statistics},
  pages={4447--4455},
  year={2024},
  organization={PMLR}
}

@inproceedings{chen2023area,
  title={Area: adaptive reweighting via effective area for long-tailed classification},
  author={Chen, Xiaohua and Zhou, Yucan and Wu, Dayan and Yang, Chule and Li, Bo and Hu, Qinghua and Wang, Weiping},
  booktitle={Proceedings of the IEEE/CVF International Conference on Computer Vision},
  pages={19277--19287},
  year={2023}
}

@inproceedings{avrahami2025stable,
  title={Stable flow: Vital layers for training-free image editing},
  author={Avrahami, Omri and Patashnik, Or and Fried, Ohad and Nemchinov, Egor and Aberman, Kfir and Lischinski, Dani and Cohen-Or, Daniel},
  booktitle={Proceedings of the Computer Vision and Pattern Recognition Conference},
  pages={7877--7888},
  year={2025}
}

@inproceedings{kipf2017semi,
  title={Semi-Supervised Classification with Graph Convolutional Networks},
  author={Kipf, Thomas N and Welling, Max},
  booktitle={ICLR},
  year={2017}
}

@article{velickovic2018graph,
  title={Graph Attention Networks},
  author={Velickovic, Petar and Cucurull, Guillem and Casanova, Arantxa and Romero, Adriana and Lio, Pietro and Bengio, Yoshua},
  journal={ICLR},
  year={2018}
}

@inproceedings{xu2017scene,
  title={Scene graph generation by iterative message passing},
  author={Xu, Danfei and Zhu, Yuke and Choy, Christopher B and Fei-Fei, Li},
  booktitle={Proceedings of the IEEE conference on computer vision and pattern recognition},
  pages={5410--5419},
  year={2017}
}

@inproceedings{ji2020action,
  title={Action genome: Actions as compositions of spatio-temporal scene graphs},
  author={Ji, Jingwei and Krishna, Ranjay and Fei-Fei, Li and Niebles, Juan Carlos},
  booktitle={Proceedings of the IEEE/CVF conference on computer vision and pattern recognition},
  pages={10236--10247},
  year={2020}
}

@article{scarselli2008graph,
  title={The graph neural network model},
  author={Scarselli, Franco and Gori, Marco and Tsoi, Ah Chung and Hagenbuchner, Markus and Monfardini, Gabriele},
  journal={IEEE transactions on neural networks},
  volume={20},
  number={1},
  pages={61--80},
  year={2008},
  publisher={IEEE}
}

@article{bronstein2021geometric,
  title={Geometric deep learning: Grids, groups, graphs, geodesics, and gauges},
  author={Bronstein, Michael M and Bruna, Joan and Cohen, Taco and Veli{\v{c}}kovi{\'c}, Petar},
  journal={arXiv preprint arXiv:2104.13478},
  year={2021}
}

@article{wu2020comprehensive,
  title={A comprehensive survey on graph neural networks},
  author={Wu, Zonghan and Pan, Shirui and Chen, Fengwen and Long, Guodong and Zhang, Chengqi and Yu, Philip S},
  journal={IEEE transactions on neural networks and learning systems},
  volume={32},
  number={1},
  pages={4--24},
  year={2020},
  publisher={IEEE}
}

@article{kim2025wow,
  title={WoW-Bench: Evaluating Fine-Grained Acoustic Perception in Audio-Language Models via Marine Mammal Vocalizations},
  author={Kim, Jaeyeon and Yun, Heeseung and Woo, Sang Hoon and Yang, Chao-Han Huck and Kim, Gunhee},
  journal={arXiv preprint arXiv:2508.20976},
  year={2025}
}

@article{kim2022pure,
  title={Pure transformers are powerful graph learners},
  author={Kim, Jinwoo and Nguyen, Dat and Min, Seonwoo and Cho, Sungjun and Lee, Moontae and Lee, Honglak and Hong, Seunghoon},
  journal={Advances in Neural Information Processing Systems},
  volume={35},
  pages={14582--14595},
  year={2022}
}

\end{document}